\documentclass[traditabstract]{aa}
\usepackage[utf8]{inputenc}
\usepackage{amsmath}
\usepackage{amssymb}
\usepackage{microtype}
\usepackage{natbib}
\usepackage{graphicx}
\usepackage{txfonts}
\usepackage{color}
\usepackage{multirow}
\usepackage{xspace}
\bibpunct{(}{)}{;}{a}{}{,} 
\newcommand{\src}{\object{Swift~J0243.6$+$6124}\xspace}

\begin{document}

\title{Orbit and intrinsic spin-up of the newly discovered transient X-ray pulsar \src.}
\author{V.\,Doroshenko\inst{1}, S.\,Tsygankov\inst{2,3}, A.\,Santangelo\inst{1}}  
\institute{Institut für Astronomie und Astrophysik, Sand 1, 72076 Tübingen, Germany\and
Tuorla Observatory, Department of Physics and Astronomy, University of Turku, V\"ais\"al\"antie 20, FI-21500 Piikki\"o, Finland\and
Space Research Institute of the Russian Academy of Sciences, Profsoyuznaya Str. 84/32, Moscow 117997, Russia}

\abstract{We present the orbital solution for the newly discovered transient Be
X-ray binary \src based on the data from gamma-ray burst monitor onboard
\emph{Fermi} obtained during the Oct 2017 outburst. We model the Doppler
induced and intrinsic spin variations of the neutron star assuming that the
later is driven by accretion torque and discuss the implications of the
observed spin variations for the parameters of the neutron star and the binary.
In particular we conclude that the neutron star must be strongly
magnetized, and estimate the distance to the source at $\sim5$\,kpc.}

\authorrunning{V. Doroshenko et al.}
\maketitle

\section{Introduction}
The transient X-ray source \src was first detected by \emph{Swift}/BAT on Oct
3, 2017 \citep{discovery_atel}. Pulsations with period $\sim9.86$\,s detected
by \emph{Swift}/XRT and \emph{Fermi}/GBM \citep{gbm_pulse}, together with the
transient behavior and tentative optical counterpart classification
\citep{opt_counterpart} suggest that it is a new Galactic Be X-ray transient.

Indeed, the follow-up \emph{NuSTAR} observations on Oct 5, 2017 \citep{nustar_atel}
revealed a cutoff power-law spectrum typical for Be transients with flux of
$\sim8.7\times10^{-9}$\,erg\,cm$^{-2}$\,s$^{-1}$. No significant spectral features, such as
a cyclotron line \citep{nustar_atel}, could be identified in broadband spectrum of the source, so no
estimate of the magnetic field strength of the neutron star could be obtained.

The outburst reached peak flux of $\sim9$ Crab level and is still
ongoing. The spin evolution of the source is being monitored by
\emph{Fermi}/GBM and appears to be mostly driven by Doppler induced variations
due to the orbital modulation. Here we report the first orbital solution for
the system based on the GBM data, and briefly discuss implications of the
observed intrinsic spin variations for the basic parameters of the system.

\section{Data analysis and results} 
\begin{figure*}[!ht]
        \centering
        \includegraphics[width=0.33\textwidth]{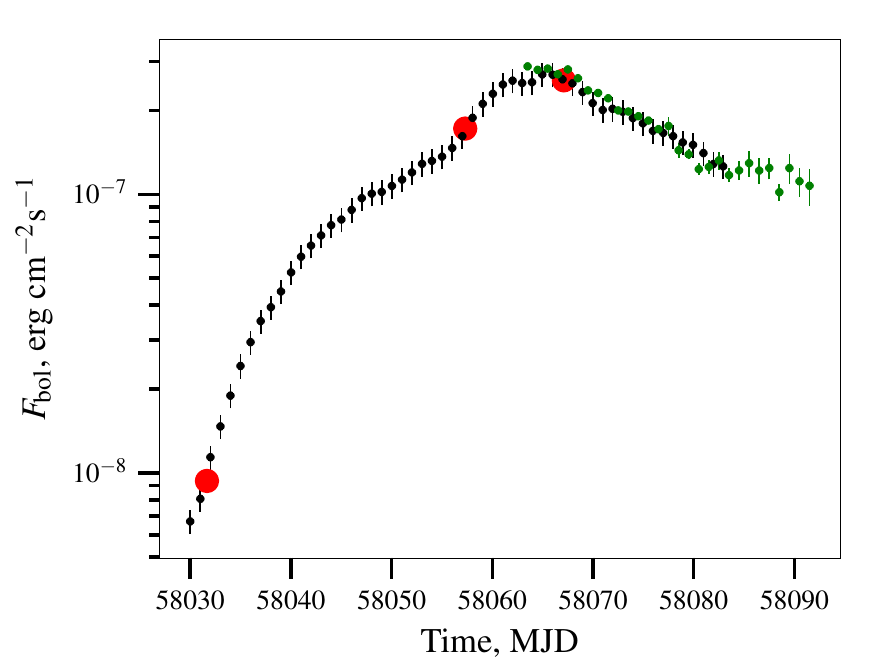}
        \includegraphics[width=0.33\textwidth]{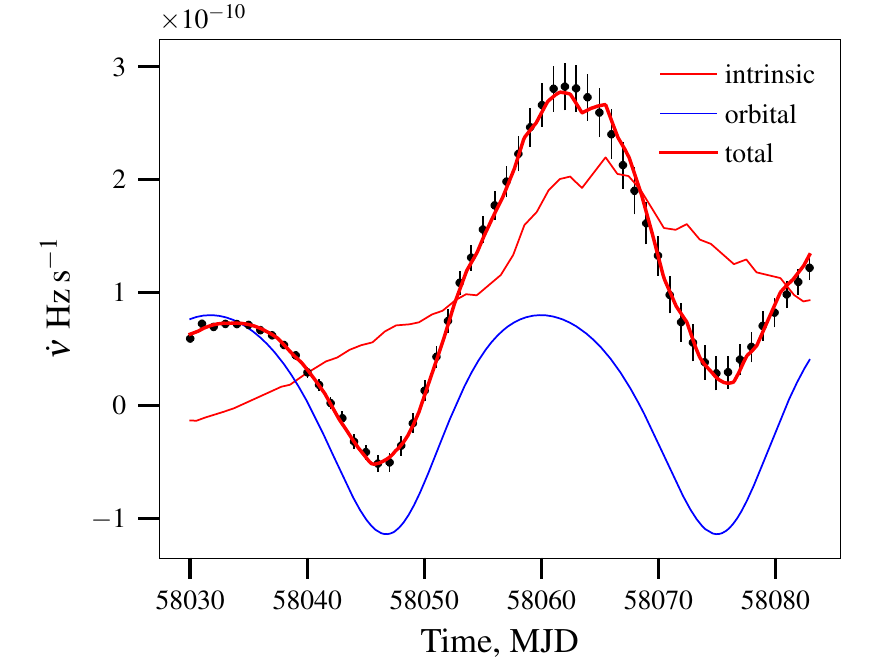}
        \includegraphics[width=0.33\textwidth]{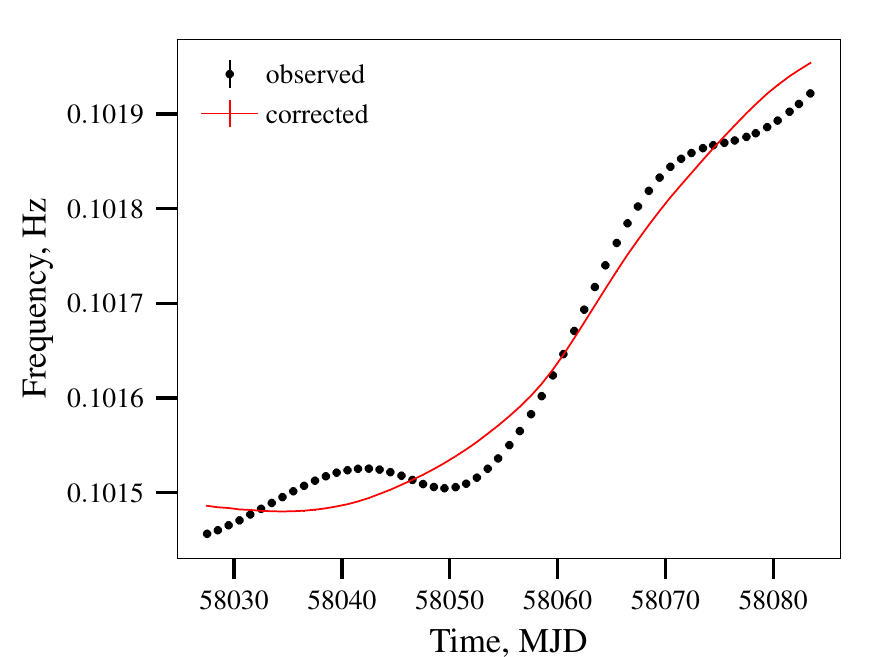}
        \caption{
        \emph{Left panel:} Bolometric lightcurve obtained by scaling
        the \emph{Swift}/BAT 15-50\,keV countrate (black error bars) to match
        the broadband flux observed by \emph{NuSTAR} (red circles). The flux derived
        from \emph{MAXI} spectra is also shown for reference (green circles).
        \emph{Middle panel:} The observed spin frequency period
        derivative reconstructed as described in the text (points) and used to
        determine the orbital parameters of the system. The best-fit model for
        the intrinsic (thin red line), the orbital-induced (thin blue line),
        and total (thick red) frequency derivative components are also shown.
        \emph{Right panel:} The spin frequencies as reported by GBM
        pulsar project (black) and orbital motion corrected using the best-fit
        ephemeris (red).}
        \label{fig:fit}
\end{figure*}

The analysis presented below is based on the spin history of the source
provided by \emph{Fermi}/GBM pulsar project
\footnote{https://gammaray.nsstc.nasa.gov/gbm/science/pulsars/lightcurves
/swiftj0243.html} from MJD~58027 to MJD~58084. Already a visual inspection of
the spin evolution (see Fig.~\ref{fig:fit}) suggests that despite the apparently high accretion rate it is
modulated by orbital motion rather than intrinsic spin-up of the pulsar. 
Still, the intrinsic spin-up is important and it is essential to model it
accurately in order to recover the orbital modulation of the spin frequency and thus orbital parameters of the binary. 

To obtain the orbital solution we used initially the same technique as
\cite{Tsyg_smc, sugizaki17}. We found, however, that for \src it yields
unsatisfactory results due to the large model systematic discussed in
\cite{Tsyg_smc}. Indeed, the observed frequency
at each moment is obtained by integration of the intrinsic spin-up rate
predicted by some torque model, which inevitably depends on the accretion rate.
The uncertainty of the observed flux translates thus to a systematic uncertainty in
predicted frequency which accumulates over time. In case of \src it appeared
excessively large which lead to overestimation of the uncertainties for the
orbital parameters. To overcome this issue, we implemented here a more direct
approach, which does not involve integration of the accretion rate.
In particular, we fit the instantaneous spin frequency derivative
rather than the observed spin frequency (i.e. similarly to \citealt{Sanna17}).
The spin-up rate and its uncertainty can be estimated directly from
the comparison of the frequencies measured in consequent time intervals
(propagating the uncertanties). The estimated frequency derivatives are then ascribed to the midpoint between respective time intervals.
The results are presented in Fig.~\ref{fig:fit} where additional
model systematic accounting for the uncertainty in the accretion rate as discussed below is added in quadrature.

The observed frequency variations are caused by a combination of Doppler shifts
due to the orbital motion and the intrinsic spin-up of the neutron star. The
radial velocity and thus the frequency change rate due to the orbital motion of
the pulsar can be unambiguously calculated for any assumed orbit by solving the
Kepler's equation which we do numerically following the equations in
\cite{Hilditch01}. On the other hand, the intrinsic spin-up rate of the
neutron star is expected to be a function of accretion rate and can be
calculated assuming some model for the accretion torque. In particular, we
assume that the intrinsic spin evolution is driven by the accretion torque
exerted by a thin threaded accretion disc, and is described with the model by
\cite{Ghosh79}. The model parameters are the mass, the radius and the magnetic
field of the neutron star, and the accretion rate. We use standard neutron star
mass and radius ($R=10$\,km, M=$1.4 M_\odot$), and consider magnetic field as a
free parameter.

The accretion luminosity can be estimated based on the observed flux.
While it is possible to use pulsed flux measured by GBM in 12--50\,keV
as a proxy, we found that non-pulsed flux measured by \emph{Swift}/BAT in
15-50\,keV range \citep{Krimm13} represents a better tracer of the bolometric
flux. To convert the observed count-rate to flux, we first cleaned the
artifact dips from the survey lighcurve and rebinned it to match the time
intervals used by the GBM. The resulting lightcurve was then scaled using the broadband fluxes 
estimated from \emph{NuSTAR} spectra
of the source observed on MJD~58031.7, 58057.3, 58067.1 assuming the same model as \cite{nustar_atel}.
To obtain the spectra we mostly followed the standard data
reduction procedures described in \emph{NuSTAR} user guide. Taking into the
account source brightness we opted, however, for
slightly larger than recommended extraction radius of $120^{\prime\prime}$.
Furthermore, for the two observations close to the peak of the outburst standard
screening criteria had to be relaxed by setting the \emph{statusexpr} parameter to ``b0000xx00xx0xx000'' 
to avoid misidentification of source counts as flickering pixels as described in \cite{2017ApJ...839..110W}.

We then estimated the bolometric flux from the spectral fit in
3-80\,keV energy range at $F_x\sim9.3\times10^{-9},1.72\times10^{-7}$, and
$2.56\times10^{-7}$\rm erg\,cm$^{-2}$\,s$^{-1}$ for the three observations.
Comparison with the contemporary BAT count rates implies then
$1.54(3)\times10^{-7}$\,erg\,cm$^{-2}$count$^{-1}$ conversion factor. The
scaled lightcurve is presented in Fig.~\ref{fig:fit} with errorbars
including the uncertainty in the conversion factor. We also verified
that estimated flux agrees with the flux measured by \emph{MAXI} monitor \citep{maxi}.
Using the daily spectra of the source
extracted using the on-demand process provided by \emph{MAXI} team\footnote{http://maxi.riken.jp/mxondem/}, and the same spectral model as above,
we calculated
the bolometric fluxes. The resulting lightcurve indeed was found to agree with the scaled
\emph{Swift}/BAT flux as shown in Fig.~\ref{fig:fit}.
Since \emph{MAXI} only observed part of the outburst, we use the
\emph{Swift}/BAT flux below to calculate the accretion luminosity for any assumed distance $d$ which we
consider a free parameter.

The other five parameters of the final model combining the intrinsic spin-up
and that induced by the orbital motion are orbital parameters of the system, i.e.
the orbital period $P_{\rm orb}$, the projection of the semimajor axis $a\sin{i}$, the eccentricity
$e$, the longitude of periastron $\omega$, and the periastron time $T_{\rm PA}$.
Statistical uncertainties in the observed flux might affect the predicted
accretion torque, so for the final fit and calculation of the uncertainties for
the best-fit parameters, we include it as additional model systematics which
is calculated by propagation of the observed flux uncertanties based on the best-fit
model obtained without inclusion of the systematics.

The best-fit results are presented in Table~\ref{tab:orpar} and
Fig.~\ref{fig:fit} where also the contribution of intrinsic accretion-driven
spin-up and of the orbital motion to the observed frequency derivative are
shown. The obtained parameters are similar to values reported by \citep{hxmt_atel} and
Fermi/GBM pulsar project. The the semimajor axis is, however, somewhat larger
for our solution, which is likely related to the difference in estimated bolometric fluxes.

Note the low (for a BeXRB) eccentricity of the orbit and comparatively
short orbital period. These, together with the high brightness of the source
suggest, that it undergoes a giant rather than normal outburst. Note
that obtained estimates for the distance to the source and magnetic field
depend on the assumed torque model. The orbital parameters are also affected to
some extent. The quoted uncertanties only reflect the statistical uncertanties
of the observed spin-up rates and fluxes, and not systematic associated with
choice of the torque model. We note, however, that this is a general problem
for X-ray pulsar timing as the intrinsic spin-up must be in any case modeled, and
using a realistic approximation for torque affecting the neutron star instead
of the more commonly used polynomial approximation is in any case more reliable.

\begin{table}
        \begin{center}
        \begin{tabular}{llll}
                Parameter & Value \\
                \hline
                $P_{orb}$, d & 28.3(2) \\
                $a\sin(i)$, lt\,s & 140(3) \\
                $e$ & 0.092(7) \\
                $\omega$, deg & -76(4) \\
                $T_{PA}$, MJD & 58019.2(4) \\
                $\chi^2/dof$ & 23.9/47 \\
                $d_{GL}$\,kpc & 6.60(5) \\
                $B_{GL}/(10^{13}\mathrm{\,G})$ & 1.08(6) \\
                \hline
        \end{tabular}
        \end{center}
        \caption{The best-fit orbital parameters of \src. Approximate values for the magnetic field strength and distance to the source assuming the \cite{Ghosh79} model are also quoted. All uncertainties are at
a $1\sigma$ confidence level and account for model systematics associated with the uncertainty of flux but not model choice.}
        \label{tab:orpar}
\end{table}
\section{Discussion and conclusions}
The intrinsic spin evolution of the neutron star is recovered as part of the
determination of the orbital parameters. It is interesting, therefore, to
discuss which implications the observed spin-up rate might have for the basic
parameters of the neutron star and the binary under various assumptions on the
accretion torque. For instance, it is possible to deduce the lower limit on
accretion rate neglecting the magnetic braking torque and assuming that the
pulsar is spun-up with maximal possible rate
\citep{Lipunov,Scott97},
$$
\dot{M}_{17}\ge0.44\nu^{1/3}\dot{\nu}/10^{-12}I_{45}M_{1.4}^{-2/3}\sim45,
$$
where $\dot{M}_{17}$ is the accretion rate in units of $10^{17}$\,g\,s$^{-1}$, $\nu$
is spin frequency, $I_{45}$ and $M_{1.4}$ are momentum of inertia and mass of the 
neutron star in units of $10^{45}$\,g\,cm$^2$ and $1.4M_\odot$ respectively.
For maximal observed intrinsic spin-up rate of 
($\dot{\nu}\sim2.2\times10^{-10}$\,Hz\,s$^{-1}$) this implies $L_x\ge
GM_{NS}\dot{M}/R\sim8.4\times10^{38}$\,erg\,s$^{-1}$ far above the Eddington limit and
approaching the levels observed in ultra-luminous X-ray sources.

The observed bolometric flux corresponding to the maximal observed spin-up rate is
$F_x\sim2.8\times10^{-7}$\,erg\,cm$^{-2}$\,s$^{-1}$, which implies a distance to
the source of $\gtrsim 5$\,kpc. We note that this limit is fairly
robust as the observed spin-up rate is only weakly affected
by the uncertainty in orbital parameters, and corresponding broadband
flux is well constrained.

\begin{figure}[!ht]
        \centering
        \includegraphics[width=0.95\columnwidth]{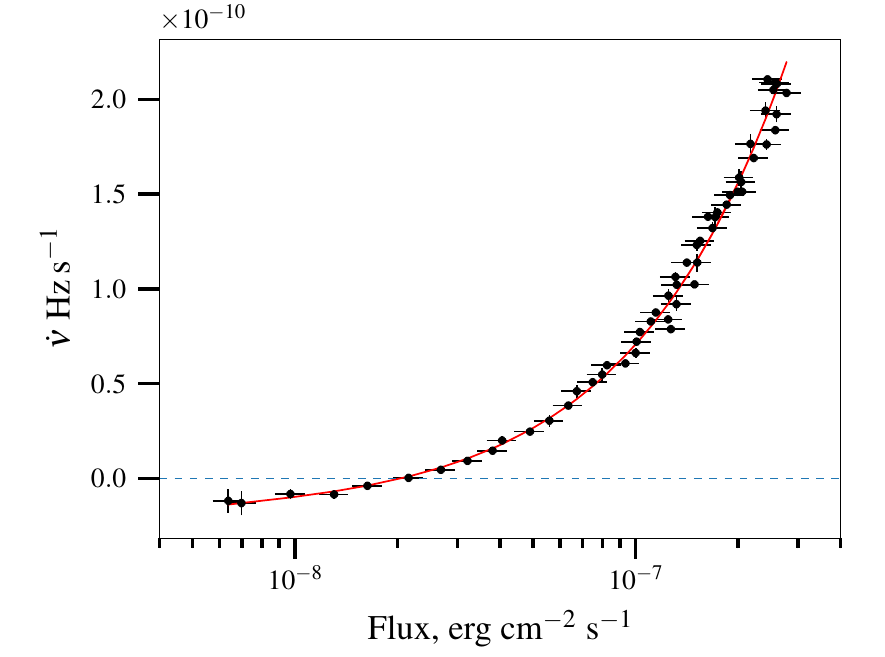}
        \caption{Intrinsic spin frequency derivative reconstructed from the corrected values and best-fit with the \cite{Ghosh79} model.}
        \label{fig:bd}
\end{figure}

The distance and the magnetic field of the neutron star are already estimated as part of
the orbit determination assuming the \cite{Ghosh79} model for torques. However,
for illustration it is useful also to compare directly the intrinsic spin-up
rate with the prediction of the model. Using the spin frequencies corrected for
orbital motion as presented in Fig.~\ref{fig:fit}, we calculated the intrinsic
spin-up rate using the same approach as above. Note that the correlation of
the spin-up rate with flux is indeed apparent as shown in
Fig.~\ref{fig:bd}. Futhermore, the source appears to spin-down at
lower fluxes, i.e. the braking torque is actually not negligible, and the limit
on the accretion rate obtained above is indeed only a lower limit.

We considered torque models by \cite{Ghosh79}, \cite{Lipunov}, \cite{Wang87},
and \cite{Parfrey16} to estimate the magnetic field of the neutron star and
the accretion rate. For all models one of the most important parameters is the
inner radius of the accretion disc $R_d=kR_{A}$, expressed as fraction of the
Alfv\'enic radius. We assumed it to be $k=0.5$ for consistency (since
$k\sim0.5$ in \cite{Ghosh79} model). Note, that this assumption is arbitrary to
some extent, however, $k$ value mostly affects the estimated magnetic field
rather than distance.
Other parameters (besides the field strength and the accretion rate depending
on distance) were kept free. In \cite{Ghosh79} and \cite{Wang87} models both
field and accretion rate are well constrained mainly due to the fact that the
inner radius is tied to the magnetosphere size and fully defines the accretion
torque (i.e. the distance and field strength are the only parameters which
affect it). The magnetosphere can be neither too small (which would imply
excessively strong spin-up at high luminosities) nor too large (as that would
inhibit accretion at low luminosities), so both the field and the accretion
rate (and hence distance) are formally well constrained.

If the coupling radius is considered a free parameter, the magnetic field
becomes correlated with $k$ and is thus poorly constrained. The \cite{Lipunov}
and \cite{Parfrey16} models also contain additional parameters characterizing
the efficiency of angular momentum transfer, so without additional
assumptions the field and distance to the source can not be constrained
simultaneously and only the lower limit on the distance discussed above holds
(because the accretion torque is the same in all models). High accretion rate
implying a distance in excess of $\sim5$\,kpc and strong field in excess of
$\sim10^{13}$\,G are, therefore, required regardless on the torque model and
model parameters.

We conclude, therefore, the source must be located further than
$\sim5$\,kpc assuming the standard neutron star parameters.  Note that this is
by a factor of two higher than distance estimated from the photometry of the
optical counterpart \citep{Bikmaev17}. The origin of this discrepancy is not
yet clear and a detailed investigation of the properties of the optical
companion is ongoing. It is important to emphasize, however, that the spin-up
rate of the neutron star is well constrained, and it is highly unlikely that we
significantly underestimate the accretion rate based on the observed bolometric
source flux, so the pulsar must indeed be further away than suggested by
\cite{Bikmaev17} unless the neutron star has a much lower momentum of inertia
than usually assumed, which is unlikely. We anticipate that this descrepancy
will be ultimately resolved with the next data release of \emph{Gaia} mission.

\begin{acknowledgements}
This work is based on spin-histories provided by \emph{Fermi}~GBM pulsar
project. This research has made use of MAXI data provided by RIKEN, JAXA and the MAXI team.

VD and AS thank the Deutsches Zentrum for Luft- und Raumfahrt (DLR)
and Deutsche Forschungsgemeinschaft (DFG) for financial support. ST
acknowledges support by the Russian Scientific Foundation grant 14-12-01287.
This research has made use of data and/or software provided by the High Energy
Astrophysics Science Archive Research Center (HEASARC), which is a service of
the Astrophysics Science Division at NASA/GSFC and the High Energy Astrophysics
Division of the Smithsonian Astrophysical Observatory.
\end{acknowledgements}

\vspace{-0.3cm}
\bibliography{biblio}   
\vspace{-0.3cm}

\end{document}